\documentclass[10pt]{article}
\usepackage{amsmath,amssymb,amsfonts,amsthm,graphicx}
\title{Superposition of two nonlinear coherent states $\frac{\pi}{2}$ out of phase and their nonclassical properties}
\author{O. Abbasi and M. K. Tavassoly
\\
\footnotesize{Atomic and Molecular Group, Faculty  of Physics, Yazd University, Yazd, Iran}
\\ \footnotesize{e-mail: mktavassoly@yazduni.ac.ir  } }

\begin{document}

\maketitle \thispagestyle{empty}
\pagestyle{headings} \markright{}

 \begin{abstract}
  Considering the concept of "{\it nonlinear coherent states}", we will study the interference effects by introducing the {\it "superposition of two classes of nonlinear coherent states"} which are $\frac{\pi}{2}$ out of phase. The formalism has then been applied to a few physical systems as "harmonious states", "$SU(1,1)$ coherent states" and "the center of mass motion of trapped ion". Finally, the nonclassical properties such as sub-Poissonian statistics, quadrature squeezing, amplitude-squared squeezing and Wigner distribution function of the superposed states have been investigated, numerically. Especially, as we will observe the Wigner functions of the superposed states take negative values in phase space, while their original components do not.
 \end{abstract}

 {\bf keyword:}
   nonlinear coherent states, superposition of coherent states, nonclassical states


\newcommand{\I}{\mathbb{I}}
\newcommand{\norm}[1]{\left\Vert#1\right\Vert}
\newcommand{\abs}[1]{\left\vert#1\right\vert}
\newcommand{\set}[1]{\left\{#1\right\}}
\newcommand{\R}{\mathbb R}
\newcommand{\C}{\mathbb C}
\newcommand{\DD}{\mathbb D}
\newcommand{\eps}{\varepsilon}
\newcommand{\To}{\longrightarrow}
\newcommand{\BX}{\mathbf{B}(X)}
\newcommand{\HH}{\mathfrak{H}}
\newcommand{\D}{\mathcal{D}}
\newcommand{\N}{\mathcal{N}}
\newcommand{\W}{\mathcal{W}}
\newcommand{\RR}{\mathcal{R}}
\newcommand{\HD}{\hat{\mathcal{H}}}

\section{Introduction}\label{sec-intro}

   Coherent state as venerable object  in quantum optics  plays an important role  in many fields of physics \cite{klauder-book, {Ali-book}}.  It is rooted in quantum physics and its relationship to classical physics.
   This notion was introduced by Glauber \cite{Glauber1} when he found them as eigenstates of the annihilation operator of the harmonic oscillator. These states were generalized in various manners. In recent years, much attention has been paid to {\it nonlinear coherent states} \cite{Matos1996} or the {\it $f$-coherent states} \cite{Manko1997} that are algebraic generalization of standard coherent states. Nonlinear coherent states have been defined as the right eigenstates of deformed annihilation operator $A=a f(\hat{n})$, where $f(\hat{n})$ is an operator-valued function of the number operator $\hat{n}=a^\dag a$. Unlike the standard coherent states, nonlinear coherent states can exhibit some nonclassical features like squeezing  \cite{Buzek1}, higher-order squeezing \cite{Truong} and sub-Poissonian statistics \cite{Mandel}. This is certainly the main reason for the increasing interest of physics researches to this field of physics, both theoretically and experimentally.
   On the other hand, {\it superposition of coherent states}
   and its extension to "nonlinear coherent states"
   (superposition of nonlinear coherent states) leads to states which
   show some of the nonclassical properties mentioned above. For instance,
   the concept of "even" and "odd" coherent states as well as nonlinear coherent states
  (as particular superpositions of coherent states and nonlinear coherent states)
   were introduced respectively in \cite{physica} and \cite{mancini}. A special quantum
   interference effect also discussed in \cite{Praxmeyer} considering the standard coherent
   states and squeezed coherent states.

  Along constructing the superposed states recently a class of states have been introduced via superposition of two "standard coherent states"  $\frac{\pi}{2}$ out of phase with respect to each other, which have some interesting properties \cite{advance}. The general form of these states are given by
 \begin{equation}\label{superposition-CSs}
   |\psi\rangle=\frac{N}{\sqrt{2}}\;(|\alpha\rangle\;+e^{i\varphi}|i\alpha\rangle),
 \end{equation}
   where $|\alpha\rangle$ is the standard coherent states obtained from the eigenvalue equation $a|\alpha\rangle=\alpha|\alpha\rangle$, $\varphi$ is a relative phase factor and $N$ is some appropriate normalization constant.\\
   Our main goal in the present paper is the extension of the latter procedure to the nonlinear coherent states. Thus, we initially construct the superposition of two sets of nonlinear coherent states whose phase difference is $\frac{\pi}{2}$, with the general form given by
 \begin{equation}\label{superposition-NCSs}
    |\psi_{f}\rangle=\frac{N_{s}}{\sqrt{2}}\;(|\alpha,f\rangle+e^{i\varphi}|i\alpha,f\rangle),
 \end{equation}
   where $|\alpha,f\rangle$ is any class of nonlinear coherent states characterized by nonlinearity function $f(n)$. Then, as an elegant feature the nonlinearity nature of the obtained states will be established manifestly by introducing the explicit {\it nonlinearity function}. By this we mean that the introduced superposed states in the present paper, and so obviously the previously introduced states in \cite{advance}, can properly classified in the "family of nonlinear coherent states" in the languages of Vogel and Man'ko  \cite{Matos1996, Manko1997}.
   In other words, any class of superposed states can be reproduced as the right eigenstate
   of a new deformed annihilation operator namely $A_s=af_s(n)$, where as we will show $f_s(n)$
   can be expressed in terms of the original nonlinearity function $f(n)$.
   Finally, after applying the proposed formalism to some particular nonlinearity functions corresponding to a few quantum physical systems we will investigate the nonclassical characters of the associated states such as sub-Poissonian behavior, quadrature squeezing and amplitude squared squeezing as a special case of higher order squeezing and Wigner distribution function.

     In view of our results, the nonclassicality nature and some new aspects of the corresponding states would be clear.
     An interesting property of the superposed states in \cite{advance} is that the two nonclassical effects
     squeezing exhibition and sub-Poissonian statistics (or antibunching phenomena) occur simultaneously, but in general there is no known connection between these two criteria.
     For some examples we may refer to the simplest representative of state superposition known as even and odd coherent states \cite{Buzek2}. While the even coherent states have squeezing (no antibunching effect), the odd coherent states have antibunching (no squeezing effects). Schleish et al studied this matter in the superposition of states $|\alpha e^{i\phi/2}\rangle$ and $|\alpha e^{-i\phi/2}\rangle$ with $\alpha \in \R$ \cite{Schleish}. According to their results these effects can be simultaneously observed only for enough large $\alpha ^2$ values.
     Generally the introduced superposition of states in the present paper also do not show these two nonclassical features simultaneously. But interestingly for some special physical systems we will observe two new nonclassical phenomena as "amplitude squared squeezing" and "sub-Poissonian statistics" occur simultaneously on some finite regions of space. In addition, when we calculated the Wigner quasi-distribution function associated to all physical systems considered in the paper, another interesting  feature is observed. According to our numerical results while each individual components of the superposition of states does not show nonclassicality behavior through verifying this criteria, all of the superposed states show this feature elegantly by taking negative values in phase space. To this end, it is worth mentioning that the presented structure has the potentiality to introduce a vast classes of nonlinear coherent states associated to different nonlinear oscillators through superposition of the original nonlinear coherent states. Clearly, knowing that $|\alpha,f=1\rangle=|\alpha\rangle$, our presented formalism readily recovers the results of \cite{advance} as a special case.
 \section{Construction of the new superposed states}
         \label{sec-n1}
 The nonlinear coherent states method is based on the deformation of standard annihilation and creation operators with an intensity dependent function $f(\hat{n})$ \cite{Matos1996,Manko1997}, according to the relations:
 \begin{equation}\label{A }
   A=af(\hat{n})=f(\hat{n}+1)a
 \end{equation}
 \begin{equation}\label{ AD}
   A^{\dag}=f^{\dag}(\hat{n})a^{\dag}=a^{\dag}f^{\dag}(\hat{n}+1),
 \end{equation}
  where $a$, $a^\dag$ and $\hat{n}=a^{\dag} a$ are bosonic annihilation, creation and number operators, respectively.
  The commutators between $A$ and $A^{\dag}$ read as:
 \begin{equation}\label{commutators}
   [A,A^{\dag}]=(\hat{n}+1)f^{\dag}(\hat{n}+1)f(\hat{n}+1)-\hat{n}f^{\dag}(\hat{n})f(\hat{n}).
 \end{equation}
   In this paper we choose $f$ to be real and nonnegative, i.e., ${f^{\dag}}(n)=f(n)$. Nonlinear coherent state should satisfy the typical eigenvalue equation
 \begin{equation}\label{eigenvalue eq}
  A|\alpha,f\rangle=\alpha|\alpha,f\rangle,\qquad       \alpha\in\C.
 \end{equation}
   The Fock space representation of these states is explicitly given by
 \begin{equation}\label{NCSs1}
   |\alpha,f\rangle=N_{f}\sum_{n=0}^\infty
   \frac {\alpha^n}{\sqrt{n!}\;[f(n)]!}\;| n  \rangle,
 \end{equation}
  where $N_{f}$ is some appropriate normalization factor determined as
 \begin{equation}\label{Nf}
   N_{f}=\left\{{\sum_{n=0}^\infty
   \frac {|\alpha|^{2n}}{n!\;{[f(n)f^\dag(n)]!}}}\right\}^{-\frac 1 2 }.
 \end{equation}
 By convention
  \begin{equation}\label{f(n)!}
    [f(n)]!\;\doteq\;f(1)f(2)...f(n),\qquad       [f(0)]!\;\doteq\;0.
  \end{equation}
  Similarly the nonlinear state $|i \alpha,f\rangle$ can be obtained by substitution $i \alpha$ instead of $\alpha$ in (\ref{NCSs1}). By using equations (\ref{superposition-NCSs}) and (\ref{NCSs1}) we are led to the explicit form of the new superposition of nonlinear coherent states as follows
 \begin{equation}\label{psi}
  |\psi_{f}\rangle=\frac{N_{s}\;N_{f}}{\sqrt{2}}\sum_{n=0}^\infty
   \frac { {\alpha^{n}\;\{1+\exp [i(\varphi\;+\;\frac{n\pi}{2})]\}} }{\sqrt{n!}\;[f(n)]!}\;| n  \rangle,
 \end{equation}
  where the normalization factor $N_{s}$ is determined from condition $\langle\psi_{f}| \psi_{f}\rangle=1$
 \begin{equation}\label{Ns}
   N_{s}=\left\{1+|N_{f}|^2\sum_{n=0}^\infty
     \frac {|\alpha|^{2\;n}\;\cos(\varphi+\frac{n\pi}{2})}{{n!}\;\;[f(n) f^\dag(n)]!}\right\}^{-\frac{1}{2}}.
 \end{equation}
  Note that in obtaining (\ref{Ns}) we have used the inner product of two nonlinear coherent states $\frac{\pi}{2}$ out of phase which can be easily written as
  \begin{equation}\label{inner product}
    \langle \alpha,f | i\alpha, f\rangle =|N_{f}|^2\sum_{n=0}^\infty
     \frac {i^n \;|\alpha|^{2\;n}}{{n!}\;[f(n)f^\dag(n)]!}.
 \end{equation}
 In fact setting $f(n)=1$ in (\ref{psi}) and (\ref{Ns}) one readily recovers
 the recently introduced superposition of the standard coherent states in \cite{advance}.\\
   As stated in \cite{advance} the introduced states obtained from the superposition of two
   standard coherent states $\frac \pi 2$ out of phase are
   the eigenstates of the fourth power of bosonic annihilation operator, $a^4$.
   This situation also holds for our constructed states in (\ref{psi}) by the expression
   $A^{4}|\psi_{f}\rangle = \alpha^{4} |\psi_{f}\rangle$.\\
  We attempt now to demonstrate that the constructed state in (\ref{psi})
  can also be classified in the nonlinear coherent states category with a special type of nonlinearity function ${f_{s}}(n)$. It is well-known that there exists a simple relation between the expansion coefficients $C_{n}$'s of the nonlinear coherent states with the corresponding nonlinearity function $f(n)$ as follows \cite{Manko1997}
 \begin{equation}\label{f(n)}
   f(n)=\frac{C_{n-1}}{\sqrt{n}C_{n}}.
 \end{equation}
   This relation helps one to recognize the nonlinearity nature of a generalized coherent state.
   Thus the nonlinearity function corresponding to the introduced states in (\ref{psi}) is readily obtained in terms of the original function $f(n)$ as
 \begin{equation}\label{fs(n)}
   f_{s}(n)=\frac{1-\exp{[i(\varphi+\frac{(n+1)\pi}{2})]}}{1+\exp{[i(\varphi+\frac{n\pi}{2})]}}\;f(n).
  \end{equation}
 The subscript "s" in this case and others which follow indicates to the "superposition" of states.
 Recall that there are many generalized coherent states that can be viewed as the nonlinear coherent states with known nonlinearity functions \cite{construction,dual}. Therefore, our presentation permits one that by substitution $f(n)$ corresponding to any nonlinear oscillator algebra in (\ref{fs(n)}) $f_{s}(n)$ associated to the special superposed state $|\psi_{f}\rangle$ in (\ref{psi}) may be found.
 Thus quite analogues to (\ref{eigenvalue eq}) it is immediately found that one may reproduce the superposition of states $|\psi_{f}\rangle\equiv |\alpha, f_s\rangle$ via the eigenvalue equation
  \begin{equation}\label{fscs}
      A_s  |\alpha, f_s\rangle = \alpha |\alpha, f_s\rangle
  \end{equation}
  where $A_s=af_s(n)$.
 It is worth mentioning that putting $f(n)=1$ in (\ref{fs(n)}) leads one to the nonlinearity function associated to the superposed state of two standard coherent states $\frac{\pi}{2}$ out of phase recently introduced  in \cite{advance}, the point that has not reported there.

 \section{Nonclassical properties of  superposed state}  \label{sec-n2}

 Nonclassical properties may be investigated through studying photon statistics, quadrature squeezing and amplitude squared squeezing. So we will express these quantities regarding the explicit expansion of superposed states in (\ref{psi}).

 \subsection{Photon statistics }

  The probability of finding $n$ photons in the state $|\psi_{f}\rangle $ is  given by:
  \begin{eqnarray}\label{p(n)}
   P(n)&=&{|\langle n|\psi_{f}\rangle|}^2\nonumber \\
      &=&\frac{|N_{s}|^2\;|N_{f}|^2\;  |\alpha|^{2\;n} }{{n!} \;[f(n)f^\dag(n)]!}\;\left[1+\cos(\varphi+\frac{n \pi}{2})\right].
  \end{eqnarray}
  Note that for original nonlinear coherent states ($|\alpha,f\rangle$ or $|i\alpha,f\rangle$) the oscillatory photon count probability (as well as other nonclassical properties) depends on the form of nonlinearity function. But this nonclassicality feature is the natural signature of our introduced states due to the existence of the term $1+\cos(\varphi+\frac{n\pi}{2})$ in the photon count probability in (\ref{p(n)}). Depending on the chosen function $f(n)$ the distribution P(n)  varies versus relative phase factor $\varphi$ and number of photons $n$.\\
 For attaining more information about photon number distribution we will deal with second order correlation parameter defined by \cite{Glauber2}
\begin{equation}\label{g2(0)2}
   g^{2}(0)= \frac{\langle{a^\dag}^2\;a^2\rangle}{\langle{a^\dag}\;a\rangle^2}.
\end{equation}
 Second order correlation function can be measured by a set of two detectors \cite{mandel}, for instance the standard Hanbury-Brown-Twiss coincidence arrangement. Considering the introduced states in (\ref{psi}), the expectation values we need in calculating (\ref{g2(0)2}) can be easily obtained as follows

\begin{equation}\label{n}
     \langle a^\dag a\rangle= |N_{s}|^2\;|N_{f}|^2\sum_{n=0}^\infty
     \frac {|\alpha|^{2(n+1)}\;\left[1-\sin(\varphi+\frac{n\pi}{2})\right]}{{n!}\;\;[f(n+1)f^\dag(n+1)]!},
\end{equation}
  \begin{equation}\label{n2}
    \langle {a^\dag}^2 a^2\rangle=|N_{s}|^2\;|N_{f}|^2\sum_{n=0}^\infty
     \frac {|\alpha|^{2n}\;n(n-1)\left[1+\cos(\varphi+\frac{n\pi}{2})\right]}{{n!}\;\;[f(n)f^\dag(n)]!}.
 \end{equation}

  In fact a state exhibits super-Poissonian, Poissonian and sub-Poissonian respectively
  if $g^{2}(0)>1$, $g^{2}(0)=1$ or $g^{2}(0)<1$, depending on the particular form of $f(n)$.
 \subsection{Quadrature squeezing }
   In order to study quadrature squeezing we consider the following hermitian quadrature operators
  \begin{equation}\label{x y}
      x=\frac{a+a^\dag}{2},\qquad   y=\frac{a-a^\dag}{2i}.
  \end{equation}
  Then the following uncertainty relation holds\\
 \begin{equation}\label{uncrtainty x y}
    \langle(\Delta x)^2\rangle\;\langle(\Delta y)^2\rangle\;\geq\frac{1}{16},
 \end{equation}
   where $\langle(\Delta x_{i})^2\rangle = \langle x_{i}^2 \rangle - {\langle x_{i} \rangle} ^2$, $x_{i}=x$ or $y$. A state is squeezed if any of the following conditions holds:
 \begin{equation}\label{delta xi2}
    \langle(\Delta x)^2\rangle\leq\frac{1}{4} \qquad or \qquad \langle(\Delta y)^2\rangle\leq\frac{1}{4}.
 \end{equation}

  Now by using equations (\ref{x y}) and (\ref{delta xi2}) the squeezing conditions lead to the following inequalities:
 \begin{equation}\label{I1}
    I_{1}= \langle a^{2}\rangle+\langle{a^\dag}^{2}\rangle-{\langle a\rangle}^2-{\langle a^\dag\rangle}^2-2\;\langle a\rangle \;\langle a^\dag\rangle+2\;\langle a^\dag a\rangle\leq0,
 \end{equation}
  or
 \begin{equation}\label{I2}
     I_{2}= -\langle a^{2}\rangle-\langle{a^\dag}^{2}\rangle+{\langle a\rangle}^2+{\langle a^\dag\rangle}^2-2\;\langle a\rangle \;\langle a^\dag\rangle+2\;\langle a^\dag a\rangle\leq0,
 \end{equation}
    respectively for $x$ and $y$ quadratures. The mean values in (\ref{I1}) and (\ref{I2}) should be evaluated with respect to $|\psi_{f}\rangle $, i.e.,
 \begin{equation}\label{a}
      \langle a\rangle=\frac{|N_{s}|^2\;|N_{f}|^2}{2}\sum_{n=0}^\infty
     \frac {\alpha^{n+1}\;{{\alpha^\ast}^{n}\;\left\{1+i+\exp\left[-i(\varphi+\frac{n\pi}{2})\right]+\exp \left[i(\varphi+\frac{(n+1)\pi}{2})\right]\right\}}}{{n!}\;\;[f^\dag(n)]!\;\;[f(n+1)]!},
 \end{equation}
 \begin{equation}\label{a2}
     \langle a^2\rangle=|N_{s}|^2\;|N_{f}|^2\sum_{n=0}^\infty
    \frac {i\;\alpha^{n+2}\;{{\alpha^\ast}^{n}}\;\sin(\varphi+\frac{n\pi}{2})}{{n!}\;\;[f^\dag(n)]!\;\;[f(n+2)]!}.
 \end{equation}
   It is worth mentioning that the mean values of $ \langle a^\dag\rangle$ and $\langle{a^\dag}^{2}\rangle$ are obtained by taking
   the complex conjugates of $ \langle a\rangle$ and $\langle a^{2}\rangle$,  respectively. Also recall that we calculated the term $\langle a^{\dag}a \rangle$ in (\ref{n}).
\subsection{Amplitude-squared squeezing }
  In order to investigate the amplitude-squared squeezing effect the following two hermitian operators have been introduced
 \begin{equation}\label{X Y}
    X=\frac{a^2+{a^\dag}^2}{2},\qquad         Y=\frac{a^2-{a^\dag}^2}{2i}.
 \end{equation}
  In fact $X$ and $Y$ are the operators corresponding to the real and imaginary parts of the square of the complex amplitude of the electromagnetic field. Heisenberg uncertainty relation of these two conjugate operators is written as
 \begin{equation}\label{uncrtainty X Y}
   \langle(\Delta X)^2\rangle\;\langle(\Delta Y)^2\rangle\;\geq \frac{1}{4}\;|\langle [X,Y] \rangle|^2.
 \end{equation}
  It follows that $|\psi_{f}\rangle $ will exhibit amplitude-squared squeezing if
 \begin{equation}\label{deltaXi2}
   \langle(\Delta X)^2\rangle\leq\frac{1}{2}\;|\langle [X,Y] \rangle|,  \qquad or \qquad \langle(\Delta Y)^2\rangle\leq\frac{1}{2}\;|\langle [X,Y] \rangle|.
 \end{equation}
    By using equations (\ref{X Y}) and (\ref{deltaXi2}) we can obtain new inequalities:
 \begin{eqnarray}\label{I3}
  I_{3}&=& \frac{1}{4}\;\Big(\langle a^{4}\rangle+\langle{a^\dag}^{4}\rangle+\langle {a^\dag}^2 a^2\rangle+\langle a^2 {a^\dag}^2\rangle-
  {\langle a^2\rangle}^2-{\langle {a^\dag}^2\rangle}^2\nonumber\\
  &-& 2\;\langle a^2\rangle \;\langle {a^\dag}^2\rangle\Big)-\;\langle a^\dag a\rangle-\frac{1}{2}\leq0,
 \end{eqnarray}
    or
 \begin{eqnarray}\label{I4}
    I_{4}&=& \frac{1}{4}\;\Big(-\langle a^{4}\rangle-\langle{a^\dag}^{4}\rangle+\langle {a^\dag}^2 a^2\rangle+\langle a^2 {a^\dag}^2\rangle+{\langle a^2\rangle}^2+{\langle {a^\dag}^2\rangle}^2\nonumber\\
    &-&2\;\langle a^2\rangle \;\langle {a^\dag}^2\rangle\Big)-\;\langle a^\dag a\rangle-\frac{1}{2}\leq0,
 \end{eqnarray}
  showing amplitude squared squeezing in $X$ or $Y$, respectively.
  Thus the following mean values are also needed for our next numerical results
 \begin{equation}\label{a4}
      \langle a^4\rangle=|N_{s}|^2\;|N_{f}|^2\sum_{n=0}^\infty
     \frac {\;\alpha^{n+4}\;{{\alpha^\ast}^{n}}\;\left[1+\cos(\varphi+\frac{n\pi}{2})\right]}{{n!}\;\;[f^\dag(n)]!\;\;[f(n+4)]!},
    \end{equation}
 \begin{equation}\label{a2ad2}
      \langle {a^2a^\dag}^2 \rangle=\frac{|N_{s}|^2\;|N_{f}|^2}{2}\sum_{n=0}^\infty
    \frac {|\alpha|^{2n}\;(n+1)(n+2)\left[1+\cos(\varphi+\frac{n\pi}{2})\right]}{{n!}\;\;[f^\dag(n)]!}.
 \end{equation}
  The expectation values $\langle {a^\dag} a \rangle$ and $\langle {a^\dag}^2 a^2 \rangle$ were calculated in (\ref{n}) and (\ref{n2}) and the mean values of $\langle{a^\dag}^{2}\rangle$ and $ \langle {a^\dag}^4\rangle$ are obtained by taking the complex conjugates of relations (\ref{a2}) and (\ref{a4}), respectively.

 \subsection{Wigner function of introduced states}
   The Wigner distribution function \cite{wigner} as a square integrable function always does exist and it is useful for calculation of averages that are essential tasks in quantum physics. For certain quantum states, this function may get negative values in parts of the phase space, the situation which is classically impossible. So negativity of this distribution function in the phase space indicates the nonclassicality nature of the states.
    The explicit form of the Wigner function associated to $f$-deformed states can be expressed as
  \begin{eqnarray}\label{wigner}
    W_{f}(x,p)&=& \frac {2} {\pi}\; N_{f}^{2}\; e^{2|\alpha|^2} \sum_{m = 0}^\infty
     \sum_{n = 0}^\infty \frac{{\alpha^{\ast}}^n \alpha^m }{(-2)^{n+m} m!\; n!\;[f(m)]!\; [f^{\dag}(n)]!}  \\ \nonumber &\times& \frac{\partial^m}{\partial \alpha^m}\;\frac{\partial^n}{\partial {\alpha^\ast}^n}(-4|\alpha|^2)
  \end{eqnarray}

   Clearly we will set $\alpha=x+i p$ on the right hand side of the above relation in our further calculations.
   As a well known fact this function will be positive in all points of phase space, when the canonical coherent states will be considered (setting $f(n)=1$ in (\ref{wigner})).
 \section{Some physical realizations of the formalism}\label{sec-realizations}
  Now we are ready to apply the presented formalism to a few generalized coherent states with known nonlinearity functions. We will regard harmonious states (HS), Gilmore-Perelomov (GP) representation of $SU(1,1)$ group and center of mass motion of trapped ion giving rise to new classes of nonlinear coherent states. Then in each case the previously mentioned nonclassical properties of the associated states in consideration will be investigated numerically. For comparison in what follows we will give the numerical results of superposed states $|\psi_{f}\rangle$ related to each physical system including the original nonlinear coherent states $\vert \alpha,f \rangle$.

\subsection{Harmonious states}
 Harmonious states first introduced by Sudarshan \cite{Sudarshan} described by
  \begin{equation}\label{har-function}
    f(n)=\frac{1}{\sqrt n},
  \end{equation}
  defined on the unit disk on the complex plane centered at the origin. Setting (\ref{har-function}) in (\ref{psi}) for the explicit form of the superposed nonlinear states one readily arrives at
 \begin{equation}\label{har-psi}
   |\psi_{f}\rangle_{HS}=\frac{N_{s}\;N_{f}}{\sqrt{2}}\sum_{n=0}^\infty
   \alpha^{n}\;(1+i^{n} e^{i\varphi})\;| n  \rangle,
 \end{equation}
  where $N_{f}$ and $N_{s}$ may also be obtained by setting (\ref{har-function}) in (\ref{Nf}) and (\ref{Ns}), respectively. The closed form of the latter constants may be calculated as follows:
 \begin{eqnarray}\label{NfHs NsHs}
  N_{f}=(1-|\alpha|^{2})^\frac{1}{2},\qquad N_{s}=\left[1+\frac{1-|\alpha|^{2}}{1+|\alpha|^{4}}(\cos\varphi-|\alpha|^{2}\sin\varphi)\right]^{-\frac{1}{2}}.
 \end{eqnarray}
 In figure 1-a the curves of P(n) is shown for above nonlinearity function and special value of $|\alpha|^{2}=0.25$ according to (\ref{p(n)}). This figure shows that the probability is vanished for $n\gtrsim4$. Also we can see that there is one or more peak in the photon distribution for any value of $\varphi$,
 Also in figure 1-b we plotted $g^{2}(0)$ against $\alpha$ for
 different fixed values of relative phase $\varphi$.
 From this figure it is seen that $g^{2}(0)<1$ for the specific regions of $\alpha\in \R$ and all values of relative phases. This implies that $|\psi_{f}\rangle $ corresponding to (\ref{har-function}) exhibits sub-Poissonian behavior in special ranges of $\alpha$ and for all studied relative phases.
 For example when $\varphi=\pi$ and the interval $0<\alpha\lesssim 0.75$,
 $|\psi_{f}\rangle$ exhibits sub-Poissonian behavior
 while for $\varphi=\frac{3\pi}{2}$ this property is seen in the range $0<\alpha\lesssim 0.55$. As shown in figures 2-a and 2-b the curves of $I_{1}$ and $I_{2}$ corresponding to squeezing effect in $x$ and $y$  operators, indicate that there is no quadrature squeezing effects in $x$ and $y$ component for the chosen values of $\varphi$ while initial state $|\alpha,f\rangle$ possess this property in $y$ component.
 In figures 3-a and 3-b the curves of $I_{3}$ and $I_{4}$ are plotted against real amplitude $\alpha$ and for different fixed values of relative
 phase difference $\varphi$. This figures imply that for all values of relative phase
 differences, amplitude-squared squeezing effect is seen in $Y$ component for superposed state $|\psi_{f}\rangle$ as well as initial state $|\alpha,f\rangle$.\\
 The Wigner distribution function using this nonlinearity function is displayed in figures 4-a, 4-b and 4-c.
 The shape of the Wigner function for the two components of the superimposed states are round hills but at different points in phase space (see figures 4-a, 4-b).
 This situation is similar to that of the standard coherent states. Both sets are positive and have a Gaussian form.   The Wigner function for the superposed states using this nonlinearity function is presented in figure 4-c.
 As an obvious matter it is seen that while the Wigner function of the components of the superposed states are positive in all points of phase space, their superposition  shows nonclassical feature via taking negative values at some points in phase space.
 \subsection{$SU(1,1)$ coherent states}
   Next we deal with GP coherent states of $SU(1,1)$ group
   whose the number state representation read as \cite{perelomov}:
  \begin{equation}\label{suGP}
   |\alpha, \kappa\rangle^{su(1,1)}_{GP} = N_f \sum_{n=0}^\infty
   \sqrt{\frac{\Gamma(n+2\kappa)}{n!}}\alpha^n \; |n\rangle,\;\;\;
   N_f=\frac{(1-|\alpha|^{2})^\kappa}{\sqrt{\Gamma(2 \kappa)}},
  \end{equation}
  which similar to harmonious states, defined on the unit disk on the complex plane centered at the origin
  and the label $\kappa$ takes the discrete values $1/2, 1, 3/2, 2,
  \cdots$.
 The  nonlinearity function in this case may be determined by (\ref{f(n)}) as
  \begin{equation}\label{fsu}
    f_{GP}(n,\kappa)=\frac{1}{\sqrt{n+2\kappa -1}}.
  \end{equation}
  Now by using (\ref{fsu}) in (\ref{psi}) we can arrive at the following form for the superposed nonlinear states $|\psi_{f}\rangle$
 \begin{equation}\label{su-psi}
     |\psi_{f}\rangle^{su(1,1)}_{GP}=\frac{N_{s}\;N_{f}}{\sqrt{2}}\sum_{n=0}^\infty
     \frac{\alpha^{n}\;(1+ i^{n} e^{i\varphi})\sqrt{(2\kappa)_{n}}}{\sqrt{n!\;}}\;| n  \rangle,
 \end{equation}
 where $(2\kappa)_{n}$ is the Pochhammer symbol
  \begin{equation}\label{pochhammer}
    (2\kappa)_{n}=\frac{\Gamma(2\kappa+n)}{\Gamma(2\kappa)}.
  \end{equation}
  Therefore, using (\ref{Ns}) with (\ref{fsu}) the closed form of the normalization factor $N_{s}$ may also be determined as
 \begin{equation}\label{Nssu}
     N_{s}=\left\{1+\frac {(1-|\alpha|^{2})^{4\kappa}}{\Gamma(2 \kappa)}\;
     \Re\left[e^{i\varphi}(1-i|\alpha|^{2})^{-2\kappa}\right]\right\}^{-\frac{1}{2}}
 \end{equation}
  where  $\Re(x)$ is the real part of $x$.\\
  Photon count probability of the introduced states in (\ref{psi}) with GP nonlinearity function was illustrated
  in figure 5-a for fixed $|\alpha|^{2}=0.25$ and $\kappa=\frac{3}{2}$. From the figure it is seen that the probability is vanished for $n\gtrsim 15$ and there is more than one peak for a fixed $\varphi$ similar to superposition of two standard coherent state \cite{advance}. Also sub-Poissonian behavior of the corresponding superposed state for different phase factors can be observed in figure 5-b.
  For all considered  relative phase differences i.e. $0, \pi\; $and$\;  \frac{3\pi}{2}$, the superposed state have sub-Poissonian behavior for special ranges of $\alpha \in \R$ that it is grater than the sub-Poissonian region corresponding to $|\alpha,f\rangle$.
  We can see that there is no quadrature squeezing in $x$  and $y$ components (figures 6-a and 6-b). Also amplitude-squared squeezing is not seen in $X$ and $Y$ component for all ranges of $\alpha$ and all considered  phases differences $\varphi$ (figures 7-a and 7-b). While the initial state $|\alpha,f\rangle$ shows quadrature squeezing in restricted range of $\alpha$ in both quadratures (but in different intervals) and amplitude-squared squeezing in $Y$ component for all $\alpha$ values.\\
  The Wigner distribution function for the present nonlinearity function  is plotted in figures 8-a, 8-b and 8-c.
  Again it is seen that the Wigner function for the two components of the superposed states are round hills but at different points in phase space (see figures 8-a, 8-b),
  the situation that is similar to that of the standard coherent states which is positive with a Gaussian form.
  The Wigner function for the superposed states using this nonlinearity function is presented in figure 8-c.
  From the figures it is seen that while the Wigner function of the components of the superposed states are positive in all phase space, their superposition shows nonclassical feature, i.e. it takes negative values at some points in phase space.

\subsection{Center of mass motion of trapped ion}
  In this section we consider the nonlinearity function that is useful in the description of the motion of a trapped ion \cite{Matos1996}:
 \begin{equation}\label{trapped-function}
   f(n)=L_{n}^{1}(\eta^2)[(n+1)\;L_{n}^{0}(\eta^2)]^{-1},
 \end{equation}
  where $L_{n}^{m}(x)$ are generalized Laguerre polynomials and $\eta$ is known as the Lamb-Dicke parameter.
  The nonlinearity also depends on the magnitude of $\eta$.
  Investigation of nonclassical features of $|\psi_{f}\rangle $ for fixed $\eta=0.2$ shows that the behavior of distribution function P(n) for fixed $|\alpha|^{2}=5.29$ is oscillatory (we followed our calculations with 50 terms in the related summations for the complicated case in (\ref{trapped-function})). Altogether, in this case P(n) vanishes  for $n\gtrsim 10$ (figure 9-a). Superposed state with this nonlinearity function, exhibits sub-Poissonian  behavior for $\varphi=\pi$ similar to original nonlinear coherent state $|\alpha,f\rangle$. For other relative phase factors, the sub-Poissonian behavior exists in a finite range of $\alpha \in \R$ (figure 9-b). Also as it is seen from figures 10-a and 10-b, there is no quadrature squeezing in $x$  and $y$ components . There is one exception in figure (10-a) which is related to the initial state $|\alpha,f\rangle$ showing squeezing in $x$ quadrature. Although the amplitude-squared squeezing region in $Y$ component has been occurred for $\varphi=0$ in $\alpha\gtrsim 1.35$, for $\varphi=\frac{\pi}{4}$ in $\alpha\gtrsim 1.1$, for $\varphi=\frac{\pi}{2}$ in $\alpha\gtrsim 0.8$  and $\varphi=\frac{3\pi}{2}$ in $\alpha\gtrsim 1.8$. In figure 11-a we can see that amplitude-squared squeezing  occurs in $X$ component only for initial state $|\alpha,f\rangle$.
  While the amplitude-squared squeezing in $Y$ component is not occurred for $|\alpha,f\rangle$, the superposed state will possess this property (figure 11-b).
  The Wigner distribution function for the latter nonlinearity function  is plotted in figures 12-a, 12-b and 12-c.
  It is observed that the Wigner function for the two components of the superposed states are round hills at different points in phase space (see figures 12-a, 12-b),
  the situation that is similar to the standard coherent states which is positive with a Gaussian form.
  The Wigner distribution function for the superpose states using this nonlinearity function is presented in figure  12-c. From the figures it is observed  that the Wigner function of the components of the superposed states are positive in all phase space, while their superposed state shows nonclassical feature, i.e., it takes negative values at some points in phase space.
  
 \section{Summary and Conclusion}
    We have constructed the superposition of two nonlinear coherent states which are $\frac{\pi}{2}$ out of phase denoted by $|\psi_{f}\rangle$ in (\ref{psi}) in a general framework. The nonlinearity nature of the introduced states was demonstrated explicitly in (\ref{fs(n)}). Therefore, unlike the "even and odd coherent states" and "even and odd nonlinear coherent states" which do not belong to the family of nonlinear coherent states, the superposed states in the present paper and obviously the special case of $f=1$ in \cite{advance} can strictly be called nonlinear coherent states.
    A special feature of our states compared with the superimposed states of \cite{advance} is that the nonclassical properties of our states can be controlled by the nonlinearity function $f(n)$.
    The oscillatory behavior of the introduced states has been justified using the photon count probability according to \ref{p(n)} which  illustrates the nonclassicality of the superposed states.
    Then the presented approach has been applied to a few physical systems, so that the nonlinear coherent state  classes in quantum optics have been considerably enlarged.
    In addition, second order correlation function, squeezing and amplitude squared squeezing for constructed states are studied.  All of the mentioned nonclassical properties  depend on the relative phase $\varphi$ as well as the form of nonlinearity function.
    The sub-Poissonian behavior is a general property of all the considered physical systems for all relative phases and all nonlinearity functions. As an interesting point in quadrature squeezing, it is seen that for all considered $f(n)$'s, there is no squeezing effect in $x$ and $y$ components. But this nonclassical effect can be observed in quadrature $x$ or $y$ for original nonlinear coherent state $|\alpha,f\rangle$.  So one may conclude that the interference effects vanishes the first order squeezing effect.  Amplitude-squared squeezing occurs depending on the physical system in hand. For example in GP nonlinearity function there is no amplitude-squared squeezing while for HS and trapped ion system, this phenomena is occurred in $Y$ component, respectively in all range and finite range of $\alpha$.\\
    Generally, from the previous literature there is no known connection between the squeezing and sub-Poissonian effects for the superposed states. Moreover our results show that for the introduced superposition of states in the present paper one can observe two new nonclassical phenomena as "amplitude squared squeezing" and "sub-Poissonian statistics" which simultaneously may be observed on some finite regions of space for some physical systems.
    We also calculated the Wigner distribution function associated to all physical systems considered in the paper.
    The special feature of these figures is that while each individual components of the superposition of states in (\ref{superposition-NCSs}) does not show the nonclassicality behavior, all of the superposed states shows this feature elegantly by taking some negative values in phase space. This is mainly arises from the quantum interference effects.\\
    The application of the proposal to other physical systems with known nonlinearity functions, for instance "photon added coherent states" and "photon subtracted coherent states" \cite{sivakumar}, and arbitrary quantum systems with known discrete spectrum like "P\"{o}schl-Teller" and "infinite well potential" \cite{posch-teller, Cruzy} and so on is a straightforward task may be done elsewhere.

 \vspace {.5 cm}

 {\bf Acknowledgement:} The authors thank the referees for useful comments which allow them to improve the paper.

 \include{thebibliography}

 \newpage
 \section{FIGURE CAPTIONS}
 \vspace {.5 cm}

  {\bf FIGs. 1-a, 1-b} The plots of photon count probability P(n) as a function of photon number $n$ and relative phase difference $\varphi$ for fixed $\alpha=0.5$  displayed in 1-a, and second-order correlation function ${g^2}(0)$  against amplitude $\alpha  \in \R$ displayed in 1-b,  for harmonious states nonlinearity function .
 \vspace {.5 cm}

  {\bf FIGs. 2-a, 2-b} The plots of the inequality $I_{1}$ corresponding to quadrature squeezing for $x$ component according to equation (22) displayed in 2-a, and the inequality $I_{2}$ corresponding to quadrature squeezing for $y$ component according to equation (23) displayed in 2-b,  against amplitude $\alpha \in \R$  for harmonious states nonlinearity function.

\vspace {.5 cm}

  {\bf FIGs. 3-a, 3-b} The plots of the inequality $I_{3}$ corresponding to amplitude-squared squeezing  for $X$ component according to equation (28) displayed in 3-a, and the inequality $I_{4}$ corresponding to amplitude-squared squeezing  for $Y$ component according to equation (29) displayed  in 3-b, against $\alpha \in \R$ for harmonious states nonlinearity function.

 \vspace {.5 cm}

   {\bf FIGs. 4-a, 4-b, 4-c} The Wigner function given by (\ref{wigner})in phase space for harmonious states nonlinearity function. Figures a, b and c are respectively plotted for the nonlinear coherent states $|\alpha, f \rangle$, $ |i \alpha, f\rangle$ and their superpositions $| \psi_f \rangle$ according to (\ref{superposition-NCSs}) setting $\alpha=0.5$ for all cases. In plotting the figure c we considered $\varphi=\pi/3$.

 \vspace {.5 cm}

 {\bf FIGs. 5-a, 5-b} The same as figure 1 except that the $SU(1,1)$ nonlinearity function is considered and $\kappa$ is fixed at  $\frac{3}{2}$.

 \vspace {.5 cm}

 {\bf FIGs. 6-a,6-b} The same as figure 2 except that the $SU(1,1)$ nonlinearity function is considered and $\kappa$ is fixed at  $\frac{3}{2}$.
 \vspace {.5 cm}

  {\bf FIGs. 7-a, 7-b} The same as figure 3 except that the $SU(1,1)$ nonlinearity function is considered and $\kappa$ is fixed at  $\frac{3}{2}$.

 \vspace {.5 cm}

 {\bf FIGs. 8-a, 8-b, 8-c} The same as figures 4, except that the $SU(1,1)$ nonlinearity function is considered. We have used $\alpha=0.5$ and  $\kappa=3/2$.

  \vspace {.5 cm}

  {\bf FIGs. 9-a, 9-b} The same as figure 1 except that the center of mass motion of trapped ion is considered for fixed $\eta=0.2$ and $\alpha=2.3$.
 \vspace {.5 cm}

 {\bf FIGs. 10a, 10-b} The same as figure 2 except that the center of mass motion of trapped ion is considered for fixed $\eta=0.2$.

 \vspace {.5 cm}

 {\bf FIGs. 11-a, 11-b} The same as figure 3 except that the center of mass motion of trapped ion is considered for fixed $\eta=0.2$.
 \vspace {.5 cm}

 {\bf FIGs. 12-a, 12-b, 12-c} The same as figures 4, except that the center of mass motion of trapped ion is considered for fixed $\eta=0.2$ and $\alpha=2.3$.
 \vspace {.5 cm}

  \vspace {.5 cm}

 \end{document}